# Stress in chromium thin films deposited by DC magnetron sputtering on grounded cupper and stainless-steel substrate holders

**M.D. Medina[1], H.I. Girón[1], K. Paucar[2], A. Talledo[1] and B.R. Pujada[1*]**

[1]Facultad de Ciencias, Universidad Nacional de Ingeniería, Lima, Perú
[2]Facultad de Ingeniería Química y Textil, Universidad Nacional de Ingeniería, Lima, Perú

[*]E-mail: bpujadab@uni.edu.pe

**Abstract.** Chromium thin films deposited on silicon substrate by DC magnetron sputtering were systematically investigated as a function of film thickness, at DC power of 50W and post-deposition annealing temperature of 200 C. Two types of grounded substrate holders, copper and stainless steel, were employed to assess substrate-dependent effects. The intrinsic stress, determined by the wafer curvature method, decreases with increasing film thickness but increases with the annealing temperature. It is observed that for thinner as-deposited chromium films, the stress showed a pronounced irreversible increase when measured immediately after deposition and after several days of aging. Films deposited on copper holders consistently exhibited higher stress values than those grown on stainless steel holders. These observations suggest that the intrinsic stress in as-deposited films is linked to the growth mechanism, while the stress increase after annealing may be related to thermally active diffusion and structural relaxation. The higher stress in films grown on copper substrate holder can likely be associated with enhanced ion bombardment due to the higher electrical conductivity of copper.

## 1. Introduction

Thin films of pure metals and metal-based composites, such as Me–DLC (metal–diamond-like carbon) and Me–N (metal–nitride) coatings, deposited by magnetron sputtering, are widely used in microelectronic, optical, and mechanical applications owing to their high hardness, good adhesion, excellent wear and corrosion resistance, and stable electrical and thermal properties. However, the performance and reliability of the films are often limited by the development of residual stresses during film growth. Excessive stress can lead to film cracking, delamination, or peeling, ultimately causing device failure. As result, understanding and controlling the origin and evolution of residual stress in thin films remains a key issue in thin-film technology [1-10].

Residual stress in thin films arises from two main sources thermal stress and intrinsic stress. Thermal stress develops after deposition during the cooling process, as a result of the mismatch in thermal expansion coefficients between the film and the substrate materials. In contrast, intrinsic stress is generated during film growth itself, and originates from non-equilibrium processes associated with nucleation, grain coalescence, atomic rearrangement, and defect incorporation in the film. In most magnetron-sputtered metallic-based films deposited at room temperature, intrinsic stress dominates, since the deposition process occurs far from thermodynamic equilibrium. The magnitude and sign of intrinsic stress are dependent of the deposition parameters, including sputtering pressure, discharge power, target–substrate distance, and substrate temperature [11–13].

The evolution of intrinsic stress in sputtered metals typically begins with compressive stress during island coalescence, shifts toward tensile stress as the film thickens, and may oscillate depending on microstructural changes. This behavior is attributed to mechanisms such as atomic peening, where energetic particles induce compressive stress by forcing atoms into interstitial sites, and grain boundary relaxation, where grain growth and defect reduction promote tensile stress at higher thicknesses [1,4,14,15].

Chromium thin films often exhibit high compressive stress under energetic growth conditions, due to the atomic peening and enhanced adatom mobility. In contrast, low-energy deposition conditions tend to promote tensile stress, which arises from the formation of porous or columnar microstructures [14,15]. Grachev *et al.* [11] investigated the evolution of intrinsic stress in sputter-deposited chromium films as a function of argon pressure and reported that the tensile stress increases with film thickness following a power-law behavior, with the exponent being strongly dependent on the sputtering pressure, as previously noted by Janssen *et al.* [15]. At low Ar pressures, stress generation is correlated with grain growth and attributed to grain-boundary shrinkage during coalescence. However, at higher pressures, increased void formation disrupts this correlation, indicating a transition to a porous microstructure and a change in the stress development mechanism.

In sputtering systems, the energy flux incident in the substrate plays a key role in determining the microstructure and stress of the growing film. This flux depends on both the plasma conditions and the electrical and thermal properties of the substrate material. While several parameter depositions have been widely studied for Cr metals, the influence of the substrate holder material, particularly on the intrinsic stress, has received little attention.

In this work, we investigate the evolution of intrinsic stress in Cr thin films deposited on silicon substrates by DC magnetron sputtering under grounded conditions. Two types of substrate holders—stainless steel and copper—were used to assess the influence of substrate holder on film growth and stress development. Special attention was given to post-deposition stress evolution and thermal annealing effects, providing insight into relaxation mechanisms and the stability of the as-deposited Cr film.

## 2. Experimental

Chromium thin films were deposited onto 2-inch Si (100) substrates by DC magnetron sputtering in a high-vacuum sputtering system (Hex-L, Korvus Technology) using a 2-inch Cr target (99.99% purity). No substrate heating or bias was applied and all deposited films were with the substrate electrically grounded. The target–substrate distance was fixed at 7.5 cm. The chamber was evacuated to a base pressure of $5 \times 10^{-6}$ mbar using a turbomolecular pump backed by a rotary vane pump. During deposition, 65 sccm (standard cubic centimeter per minute) of high-purity Ar gas was introduced through a mass flow controller to achieve a working pressure in the range of $5 \times 10^{-3}$ mbar. Two sets of films were deposited at 50 W DC power on grounded stainless steel (SS) and copper (Cu) substrate holders to assess the influence of substrate conductivity on plasma–substrate interaction and stress development.

The film thickness was systematically varied to investigate the thickness-dependent evolution of intrinsic stress. Thickness variations were achieved by adjusting the deposition time from 1 minute to 1 hour, resulting in chromium films with thicknesses ranging from approximately 17 nm to 1000 nm. The film thickness was determined from the mass difference of the silicon wafer before and after deposition, using the bulk density of chromium (7.2 g cm$^{-3}$) and the known substrate area of the silicon wafer.

The intrinsic stress in the Cr films was evaluated using the wafer curvature method with a two-beam optical system, measured before and after deposition [13]. The curvature change ($\Delta \kappa$) was converted to film stress ($\sigma$) according to Stoney's equation [16]:

$$\sigma = \frac{E_s t_s^2}{6(1 - \nu_s) t_f} \Delta \kappa$$

where $E_s$ and $\nu_s$ are the Young's modulus and Poisson's ratio of the Si substrate, $t_s$ is the substrate thickness, and $t_f$ is the film thickness. The total uncertainty in the calculated stress values was estimated to be within ±0.1 GPa, considering contributions from curvature measurement accuracy and film thickness determination.

To investigate stress relaxation mechanisms, wafer curvature measurements were conducted after deposition over time intervals from a few hours up to several days. Subsequently, the films were annealed in a Tefic Biotech furnace under argon at 200 °C for 60 min and then cooled to room temperature. Post-annealing stress measurements were performed to assess thermally activated relaxation and structural modifications. As no substrate heating was employed during deposition, the resulting stress values predominantly represent intrinsic rather than thermal components.

Atomic force microscopy (AFM) measurements were performed to examine the surface morphology and topographical features of the chromium films deposited under different sputtering conditions. The analyses were carried out using an *nGauge* AFM system (ICSPI Corp.), which allows high-resolution imaging in ambient conditions. Surface scans were obtained over representative areas of the samples to evaluate parameters such as surface roughness, grain size, and morphological uniformity.

To evaluate the presence of elements other than chromium in the deposited films, their elemental composition was analyzed using energy-dispersive X-ray spectroscopy (EDS) integrated into a ZEISS EVO 15 scanning electron microscope (SEM). The EDS measurements were carried out with a focused electron beam at an accelerating voltage of 10 keV. This energy was chosen to minimize signal contribution from the underlying silicon substrate and ensure that the analysis remained confined to the film surface.

## 3. Results

Figure 1 shows the time-dependent evolution of intrinsic stress in a chromium film deposited by DC magnetron sputtering for 5 minutes on a stainless-steel substrate holder at a power of 50 W. For this deposition time, the calculated film thickness was approximately 78 nm. From fig. 1 is observed that immediately after deposition, the film exhibits a tensile stress of about 0.55 GPa. Over time, the tensile stress gradually increases, reaching a mean value of 0.81 GPa after several days. An oscillatory behaviour in the intrinsic stress is observed during the first few days following deposition, after which the stress stabilizes at a higher value.

For thinner films (<78 nm), the oscillatory behaviour of the stress is more pronounced during the initial hours after deposition, although the variations remain within the experimental uncertainty. In contrast, for thicker films (>400 nm), the stress remains essentially constant over time, within the limits of experimental error.

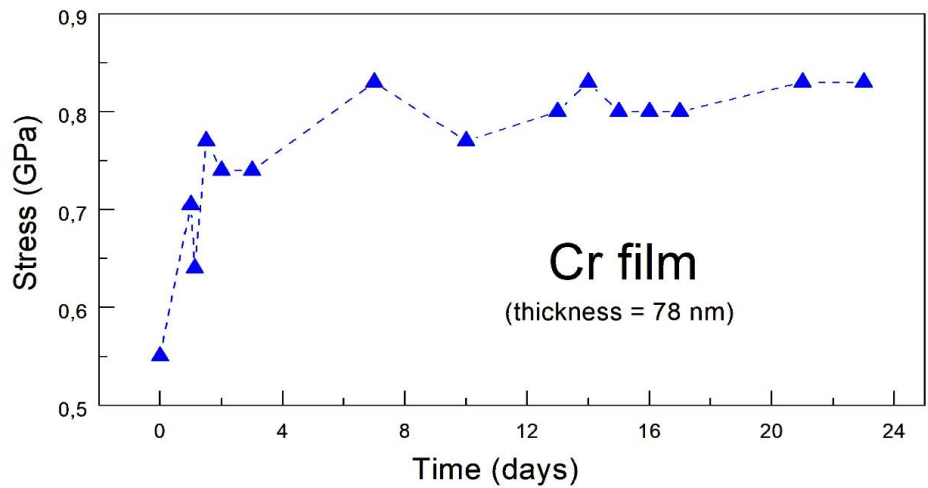

**Figure 1.** Time evolution of the intrinsic stress in 78 nm-thick Cr film deposited using 50 W DC power on a stainless-steel substrate holder. The stress was measured immediately after deposition and subsequently monitored over several days. A progressive increase in stress is observed, stabilizing after approximately seven days of aging.

A similar trend was observed for chromium films deposited on a copper substrate holder under the same DC power of 50 W. However, in this case, the maximum stress value was reached after only three days, compared to seven days for films deposited on the stainless-steel substrate holder. These findings suggest that the physical mechanism responsible for the observed stress evolution is independent of the substrate holder material, although in the case of copper, the stress reaches its maximum value more quickly than with stainless steel.

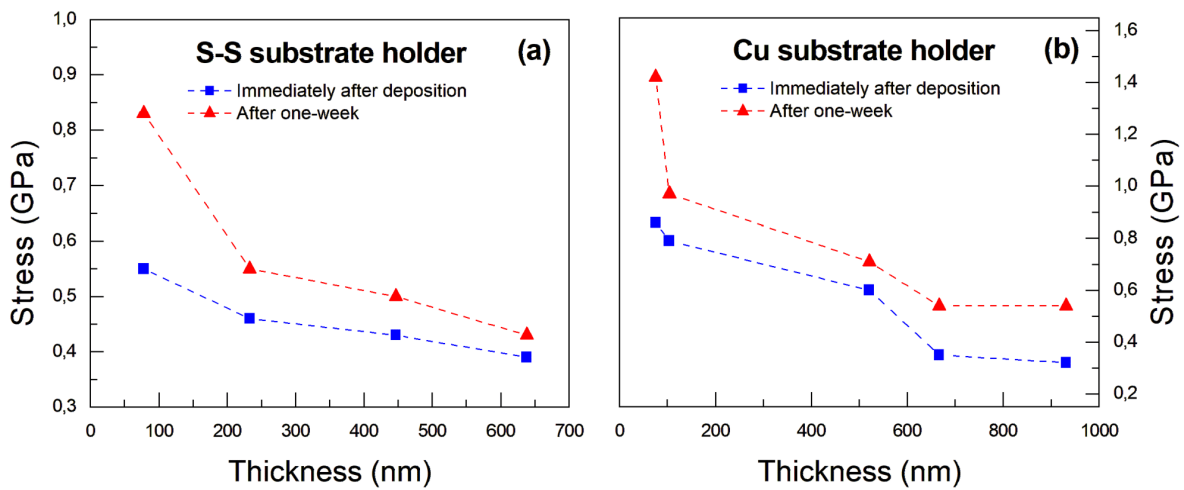

**Figure 2.** Intrinsic stress as a function of film thickness for Cr films measured immediately after deposition (blue) and one week later (red), for samples deposited using (a) a stainless-steel (S-S) substrate holder and (b) a cupper substrate holder. Dashed lines are included as guides to the eye.

Figure 2(a–b) shows the dependence of intrinsic stress on film thickness for chromium layers deposited by DC magnetron sputtering at 50 W, using stainless-steel (a) and copper (b) substrate holders. Stress measurements were performed immediately after the silicon wafers were removed from the sputtering chamber (blue) and repeated one-week post-deposition (red), enabling a direct comparison between the as-deposited stress state and its evolution under ambient conditions.

For both substrate holders, a clear inverse relationship between intrinsic stress and film thickness is observed: the stress decreases as the chromium film thickness increases. Notably, for thinner films, the stress values measured after one-week are up to 50% higher than the initial values. In contrast, for thicker films, the increase in stress over time is significantly lower, suggesting the operation of complex post-deposition relaxation mechanisms. These may include processes such as adatom diffusion, grain coalescence, and defect evolution, which are further influenced by environmental factors like the high ambient humidity in Lima (~80%).

Figure 3(c) presents the evolution of stress in Cr films as a function of film thickness, measured after annealing at 200 °C for 1 hour, using copper substrate holder. The dashed lines in each plot correspond to the stress values recorded one-week after deposition, prior to any thermal treatment. Similar trend was observed for Cr films using stainless-steel substrate holder (not shown here). In both cases, an overall increase in stress is observed following annealing. This increase is more pronounced in thinner films. The data suggest a clear dependence of post-annealing stress on film thickness, indicating that thinner Cr layers are more susceptible to changes induced by thermal treatment.

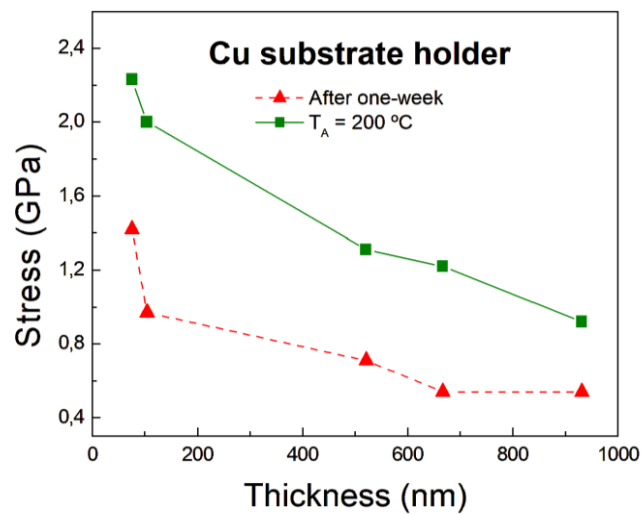

**Figure 3.** Intrinsic stress as a function of film thickness for Cr films measured one-week after deposition (dashed red line) and after annealing at $T_A$=200 °C for 1 hour (green line), for samples deposited using a copper substrate holder.

Figure 4(a–c) presents AFM surface profiles of chromium films using a copper substrate holder with thicknesses of 76 nm (a) and 931 nm (b) immediately after sputtering deposition, as well as the 931 nm film subjected to post-deposition thermal annealing at 200 °C (c). The images reveal that grain size increases with film thickness, consistent with enhanced adatom mobility and grain growth during film growth.

Notably, thermal annealing of the 931 nm film results in an apparent reduction in grain size. This phenomenon can be attributed to the densification of the film microstructure: the as-deposited films exhibit a porous, columnar morphology, whereas annealing facilitates atomic diffusion that fills intergranular voids and promotes grain boundary rearrangement. Consequently, the porosity is partially eliminated, and new, finer grains form within the original columnar grains, leading to a more compact and refined microstructure.

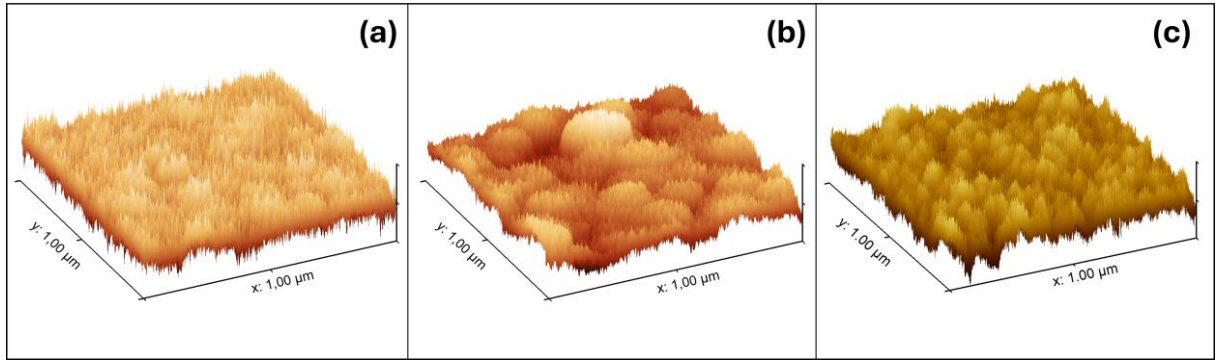

**Figure 4.** Three-dimensional AFM surface profiles of Cr films deposited using a copper substrate holder with thicknesses of (a) 76 nm, (b) 931 nm, and (c) 931 nm after annealing at 200 °C.

EDS measurements performed on Cr films using a copper substrate holder with a thickness of 521 nm revealed an oxygen content of 18 at% for the sample stored under ambient air conditions for 7 days after deposition, and 16 at% for the sample annealed at 200 °C. The relatively high oxygen concentration suggests that oxygen incorporation is not solely due to surface oxidation but may also result from moisture adsorption or water uptake during exposure to the laboratory environment, where the relative humidity is approximately 80%.

## 4. Discussion

The progressive increase in intrinsic stress with time, as shown in Fig. 1 can be attributed to post-deposition microstructural evolution processes [17]. Chromium films typically grow via the Volmer–Weber mechanism, where island coalescence and grain boundary formation generate tensile stress during film growth [15]. Following deposition, the films gradually evolve toward thermodynamic equilibrium. EDS analysis revealed an oxygen content of approximately 18 at.% for a 521 nm-thick as-deposited Cr film, suggesting substantial surface oxidation and the likely presence of adsorbed water molecules. Comparable observations were reported by Scher *et al*. [18], who demonstrated that ambient humidity promotes the formation of thin water layers on oxidized metallic surfaces, such as aluminium. The present results therefore indicate that the dominant mechanism governing the stress evolution is defect relaxation, potentially enhanced by moisture-assisted surface diffusion under the high relative humidity conditions prevalent in Lima (~80%). This moisture-driven enhancement of tensile stress over time has not, to the best of our knowledge, been previously reported for chromium films. Further investigations are warranted to elucidate the role of ambient humidity in the intrinsic stress evolution of chromium and other metallic films deposited by magnetron sputtering.

The results presented in Figs. 2 and 3 demonstrate that the intrinsic stress in the Cr films is influenced by the deposition parameters, annealing temperature, and substrate holder material. In general, the stress magnitude decreases with increasing film thickness. This trend is consistent with the well-established model in which tensile stress arises from grain boundary interactions, an effect that becomes more pronounced in the fine-grained microstructure typical of thinner films. Upon exposure to ambient air under the high-humidity conditions of Lima (~80% RH), all films exhibited a progressive increase in tensile stress. This post-deposition stress evolution was strongly dependent on film thickness, with thinner films displaying the most pronounced increase. Such behavior suggests that the tensile stress enhancement is closely related to the microstructural characteristics of the films, particularly the density and porosity of the columnar grains observed by AFM (Fig. 4). These features likely facilitate the adsorption and diffusion of moisture within the film, thereby promoting stress relaxation and subsequent tensile stress development over time.

The intrinsic stress was also found to increase with annealing temperature. This behavior can be attributed to thermally activated grain growth and defect relaxation processes, which promote microstructural densification and, consequently, higher tensile stress, consistent with previously reported results [17]. Furthermore, films deposited on copper substrate holders exhibited higher stress values than those grown on stainless steel. This difference is likely related to the higher electrical conductivity of copper, which alters the local potential distribution near the substrate during deposition, thereby enhancing ion bombardment and secondary electron emission. These effects increase the energy of the incident species, influencing film densification, stress development, and defect concentration, in accordance with the general mechanisms proposed for energetic growth conditions in metallic films [11,15]. Further systematic studies are required to gain a more detailed understanding of the influence of the grounded substrate-holder material on the intrinsic stress development in metallic thin films.

## 4. Conclusions

Chromium thin films deposited by DC magnetron sputtering exhibit tensile intrinsic stress that increases over time, primarily due to post-deposition defect relaxation and moisture-assisted surface diffusion under the high ambient humidity conditions of Lima (~80% RH). The intrinsic stress decreases with increasing film thickness but increases with annealing temperature, reflecting the effects of thermally activated grain growth and defect relaxation. Films deposited on copper substrate holders exhibit higher stress than those on stainless steel, likely as a result of enhanced ion bombardment and substrate–plasma interactions. These results demonstrate that both environmental factors and substrate-holder properties significantly influence the stress evolution in metallic thin films. Future work will aim to systematically quantify the effects of ambient humidity and substrate-holder material on intrinsic stress development.

**Acknowledgements**
We thank the Vicerrectorado de Investigación of the Universidad Nacional de Ingeniería (Perú) and the Special Formative Research Project FC-PFE-02-2023.

**References**

[1] F.A. Doljack and R.W. Hoffman, *Thin Solid Films* **12**, 71 (1972).
[2] J.A. Thornton, *J. Vac. Sci. Technol*. A **4**, 3059 (1986).
[3] W.D. Nix and B.M. Clemens, *J. Mater. Res.* **14**, 3467 (1999).
[4] F. Spaepen, Acta Mater. **48**, 31 (2000).
[5] K. Ohring, Materials Science of Thin Films, 2nd ed. (Academic Press, 2002).
[6] L.B. Freund and S. Suresh, Thin Solid Materials (Cambridge University Press, 2003).
[7] B.R. Pujada et al., J. Appl. Phys. **105**, 033502 (2009).

[8]     G.C.A.M. Janssen, et al., *Thin Solid Films* **517**, 1858 (2009).
[9]     J.E. Green, *J. Vac. Sci. Technol*., A **35**, 05C204 (2017).
[10]    G. Abadias, et al., *J. Vac. Sci. Technol. A* **36**, 020801 (2018).
[11]    S.Y. Grachev, et al., *J. Appl. Phys*. **97**, 073508 (2005).
[12]    N.Z. Calderon et al., *J. Phys. Conf. Ser*. **1558**, 012008 (2020).
[13]    S. Ponce et al., *J. Phys. Conf. Ser*. **1558**, 012009 (2020).
[14]    D.W. Hoffman, *Thin Solid Films*, **40**, 355 (1977).
[15]    G.C.A.M. Janssen, et al., *Appl. Phys. Lett*. **83**, 3287 (2003).
[16]    G.G. Stoney, *Proc. R. Soc. Lond. A* **82**, 172 (1909).
[17]    H.Z. Yu et al., *J. Appl. Phys*. **115**, 043521 (2013).
[18]    Scher et al., *ACS Appl. Mater. Interfaces*, **15**, 28716 (2023).